\documentclass[twocolumn,twoside,slac_two]{revtex4}
\usepackage{graphicx}
\usepackage{fancyhdr}
\begin{document}

\title{Blazar Variability and Evolution in the GeV Regime}

%

\author{S. Tsujimoto, J. Kushida, K. Nishijima, and K. Kodani}
\affiliation{Tokai university, Hadano city, Japan, 259-1292}

\begin{abstract}
One of the most important problem of the blazar astrophysics is to understand the physical origin of the blazar sequence. In this study, we focus on the GeV gamma-ray variability of blazars and evolution perspective we search the relation between the redshift and the variability amplitude of blazars for each blazar subclass. We analyzed the Fermi-LAT data of the TeV blazars and the bright AGNs (flux $\geq$ 4$\times10^{-9}$ cm$^{-2}$s$^{-1}$) selected from the 2LAC (the 2nd LAT AGN catalog) data base. As a result, we found a hint of the correlation between the redshift and the variability amplitude in the FSRQs. Furthermore the BL Lacs which have relatively lower peak frequency of the synchrotron radiation and relatively lower redshift, have a tendency to have a smaller variability amplitude.
\end{abstract}

\maketitle

\thispagestyle{fancy}


\section{Introduction}
The blazar is the class of the active galactic nuclei (AGN) which has the most number of extragalactic source in the very-high-energy gamma-ray regime ($E>100$ GeV)\cite{TeVCat}.
They are characterized by double-peaked nonthermal emission with spectral energy distribution (SED) in radio to gamma-ray regime.
Blazars include BL Lacertae objects (BL Lacs) and flat spectrum radio quasars (FSRQs). 
In addition BL Lacs include high-frequency peaked BL Lac objects (HBLs), intermediate-frequency peaked BL Lac objects (IBLs) and low-frequency peaked BL Lac objects (LBLs).
In the leptonic model, the low and high bump of SED
is explained by the synchrotron and the Synchrotron Self Compton (SSC) and/or External Compton (EC) radiations.
The estimation of EC in the second hump is a matter of great importance in estimating external photons.

One of the characteristic of the blazar spectra is blazar sequence.
In 1998, Fossati et al. combined three complete blazar samples\cite{Fossati1998};
The 2 Jy samples of FSRQs\cite{Padovani1992}, the radio selected 1 Jy samples of BL Lacs\cite{Kuhr1981} and the X-ray selected sample(Einstein Slew Survey) of BL Lacs \cite{Elvis1992}.
The thirty-third sources of selected sample were detected in high-energy gamma-ray regime ($E>100$ MeV) by the EGRET instrument on-board the compton gamma ray observatory (CGRO).
These sources were devided 5 bins based on the 5 GHz radio luminosity and averaged the SED of the each type of the blazars.
The made SED suggested some relationship; 
first, the synchrotron peak frequency and the bolometric luminosity have the anti-correlation.
second, the synchrotron peak frequency and the compton peak frequency have the positive-correlation.
finally, the compton dominance (the ratio of the inverse Compton to synchrotron luminosity) and the bolometric luminosity have positive-correlation.
These correlations are known as ``the blazar sequence'' in considering the blazar physics.
We aim to reveal the relation between the evolution process of AGNs and the blazar sequence based on the systematical study for many blazars .

In this study, we calculated $\sim100$ AGNs (blazars) 
to find the difference of the variability amplitude in blazar types 
and evolution of the variability amplitude (activity).
We applied the fractional variability
amplitude ($F_{\rm var}$) to calculate the variability amplitude considering the error.
The $F_{\rm var}$ is defined as Eq.(\ref{eq:Fvar}) which was given by Vaughan et al. (2003).

\noindent
\begin{equation}\label{eq:Fvar}
F_{\rm var}=\sqrt{ \frac{S^2 - \overline{\sigma_{err}^2}}{\overline{F}^2} }
\end{equation}

\noindent
Note that $S^2$ is the total variance of the light curve, 
 $\overline{\sigma_{err}^2}$ is the mean square of flux error and $\overline{F}^2$ is the square of mean flux. 
Fvar error (uncertainty) is defined as Eq.(\ref{eq:Fvarerr}) by Poutanen et al. (2008).

\noindent
\begin{equation}\label{eq:Fvarerr}
\Delta F_{\rm var}=\sqrt{F_{\rm var}^2 + err({\sigma_{{\rm {\tiny{NXS}}}}^2})}-F_{\rm var}
\end{equation}

\noindent
Where $err({\sigma_{{\rm {\tiny{NXS}}}}^2})$ is defined as Eq.(\ref{eq:Fvarerr2}) by Vaughan et al. (2003)
\begin{equation}\label{eq:Fvarerr2}
err({\sigma_{{\rm {\tiny{NXS}}}}^2}) = 
\sqrt{{\left(\sqrt{\frac{2}{N}}\frac{\overline{\sigma_{err}^2}}{F^2}\right)}^2+{\left(\sqrt{\frac{\overline{\sigma_{err}^2}}{N}}\frac{2F_{\rm var}}{F}\right)}^2}
\end{equation}

\noindent 
$F_{\rm var}$ is often used in computation variability amplitude for each spectral band\cite{MAGIC2014}.

In this study, we studied the flux variation of AGNs with the $F_{\rm var}$ on the high-energy gamma-ray regime to get the variation charactor of each class (type).

\section{DATA SELECTION}

\begin{table*}[t]
\begin{center}
\caption{Analyzed AGN list \cite{2LAC,TeVCat}}
\begin{tabular}{ccc|ccc|ccc}
\hline
\hline
\textbf{Source} & \textbf{Redshift} 
& \textbf{Class} 
& 
\textbf{Source} & \textbf{Redshift} 
& \textbf{Class} 
&
\textbf{Source} & \textbf{Redshift} 
& \textbf{Class} 
\\
\textbf{name} & \textbf{z} 
& \textbf{(type)} 
& 
\textbf{name} & \textbf{z} 
& \textbf{(type)} 
&
\textbf{name} & \textbf{z} 
& \textbf{(type)} 
\\
\hline
Messier 87 & 0.0044 &  FRI & MS 1221.8+2452 & 0.218 &  HBL & 4C +55.17 & 0.899298 &  FSRQ\\
NGC 1275 & 0.017559 &  FRI & PKS 0301-243 & 0.26 &  HBL & PKS 0823-223 & 0.91 &  IBL\\
Mrk 421 & 0.031 &  HBL & S2 0109+22 & 0.265 &  IBL & PKS 0420-01 & 0.916 &  FSRQ\\
Mrk 501 & 0.034 &  HBL & 1ES 0414+009 & 0.287 &  HBL & AO 0235+164 & 0.94 &  LBL\\
1ES 2344+514 & 0.044 &  HBL & S5 0716+714 & 0.31 &  IBL & S3 0218+357 & 0.944 &  Blazar\\
Mrk 180 & 0.045 &  HBL & OT 081 & 0.322 &  LBL & OP 313 & 0.997249 &  FSRQ\\
1ES 1959+650 & 0.048 &  HBL & 1ES 0502+675 & 0.341 &  HBL & PKS 0454-234 & 1.003 &  FSRQ\\
AP Lib & 0.049 &  LBL & PKS 1510-089 & 0.361 &  FSRQ & 4C +14.23 & 1.038 &  FSRQ\\
1ES 1727+502 & 0.055 &  HBL & 3C 66A & 0.41 &  IBL & PKS 2201+171 & 1.076 &  FSRQ\\
BL Lacertae & 0.069 &  IBL & PKS 0735+17 & 0.424 &  LBL & PKS 0426-380 & 1.111 &  LBL\\
PKS 2005-489 & 0.071 &  HBL & 4C +21.35 & 0.432 &  FSRQ & PKS B1908-201 & 1.119 &  FSRQ\\
RGB J0152+017 & 0.08 &  HBL & 1ES 0647+250 & 0.45 &  HBL & OG 050 & 1.254 &  FSRQ\\
1ES 1741+196 & 0.083 &  HBL & PG 1553+113 & 0.5 &  HBL & PKS 1551+130 & 1.30814 &  FSRQ\\
W Comae & 0.102 &  IBL & GB 1310+487 & 0.501 &  FSRQ & PKS 0244-470 & 1.385 &  FSRQ\\
1ES 1312-423 & 0.105 &  HBL & PKS 2326-502 & 0.518 &  FSRQ & PKS 2023-07 & 1.388 &  FSRQ\\
VER J0521+211 & 0.108 &  IBL & 3C 279 & 0.536 &  FSRQ & S4 1030+61 & 1.40095 &  FSRQ\\
PKS 2155-304 & 0.116 &  HBL & MG2 J071354+1934 & 0.54 &  FSRQ & PKS 0402-362 & 1.417 &  FSRQ\\
B3 2247+381 & 0.1187 &  HBL & BZQ J0850-1213 & 0.566 &  FSRQ & PKS 0250-225 & 1.419 &  FSRQ\\
RGB J0710+591 & 0.125 &  HBL & PKS 1424+240 & 0.6035 &  IBL & PKS 1454-354 & 1.424 &  FSRQ\\
H 1426+428 & 0.129 &  HBL & 4C 31.03 & 0.603 &  FSRQ & B2 1520+31 & 1.484 &  FSRQ\\
1ES 1215+303 & 0.13 &  HBL & PMN J2345-1555 & 0.621 &  FSRQ & PKS 2052-47 & 1.489 &  FSRQ\\
PKS 1717+177 & 0.137 &  LBL & PKS 1244-255 & 0.633 &  FSRQ & PKS 2227-08 & 1.55999 &  FSRQ\\
1ES 0806+524 & 0.138 &  HBL & S4 1849+67 & 0.657 &  FSRQ & TXS 1013+054 & 1.7137 &  FSRQ\\
1ES 0229+200 & 0.14 &  HBL & 4C +56.27 & 0.664 &  LBL & PKS 0215+015 & 1.721 &  FSRQ\\
1RXS J101015.9-311909 & 0.142639 &  HBL & S5 1803+784 & 0.68 &  LBL & MG1 J123931+0443 & 1.76095 &  FSRQ\\
TXS 1055+567 & 0.14333 &  IBL & Ton 599 & 0.724565 &  FSRQ & MG2 J101241+2439 & 1.805 &  FSRQ\\
3C 273 & 0.158 &  FSRQ & B2 0716+33 & 0.779 &  FSRQ & 4C +38.41 & 1.81313 &  FSRQ\\
H 2356-309 & 0.165 &  HBL & TXS 0106+612 & 0.785 &  FSRQ & PKS 0805-07 & 1.837 &  FSRQ\\
PKS 0829+046 & 0.173777 &  LBL & PKS 1622-253 & 0.786 &  FSRQ & PKS 1502+106 & 1.83928 &  FSRQ\\
1ES 1218+304 & 0.182 &  HBL & B2 2234+28A & 0.795 &  LBL & 4C 01+02 & 2.099 &  FSRQ\\
1ES 1101-232 & 0.186 &  HBL & PKS 0440-00 & 0.844 &  FSRQ & PKS 1329-049 & 2.15 &  FSRQ\\
1ES 0347-121 & 0.188 &  HBL & OC 457 & 0.859 &  FSRQ & S4 0917+44 & 2.18879 &  FSRQ\\
RBS 0413 & 0.19 &  HBL & 3C 454.3 & 0.859 &  FSRQ & PMN J1344-1723 & 2.506 &  FSRQ\\
OX 169 & 0.211 &  FSRQ & TXS 1920-211 & 0.874 &  FSRQ &  &  & \\
1ES 1011+496 & 0.212 &  HBL & PKS 0537-441 & 0.892 &  LBL &  &  & \\
\hline
\hline
\end{tabular}
\label{tab:blazarlist}
\end{center}
\end{table*}

We selected AGNs from the second LAT AGN catalog (2LAC)\cite{2LAC} and TeVCat\cite{TeVCat}.
Selection criteria were as follows,
\begin{enumerate}
\renewcommand{\labelenumi}{\Roman{enumi}} 
\item:  It was decided which subclass was belonged to (HBL or IBL or LBL or FSRQ).
\item:  It had known redshift.
\item:  flux $\geq 4\times10^{-9}$ cm$^{-2}$ s$^{-1}$ in 2LAC.\label{enu:flux}
\end{enumerate}

Table \ref{tab:blazarlist} shows the analyzed AGN list which 102 sources are included in.
Source name, redshift, and class(type) are cited from 2LAC and TeVCat.

The redshift of PKS 1424+240 was referred to Furniss(2013)\cite{Furniss2013}.
According to the TeVCat\cite{TeVCat}, the blazar class of S3 0218+357 (z=0.944) was not determined
but we used it as high redshift VHE gamma-ray emitter.

\section{DATA ANALYSIS}

\begin{figure*}[t]
\centering
\includegraphics[width=170mm]{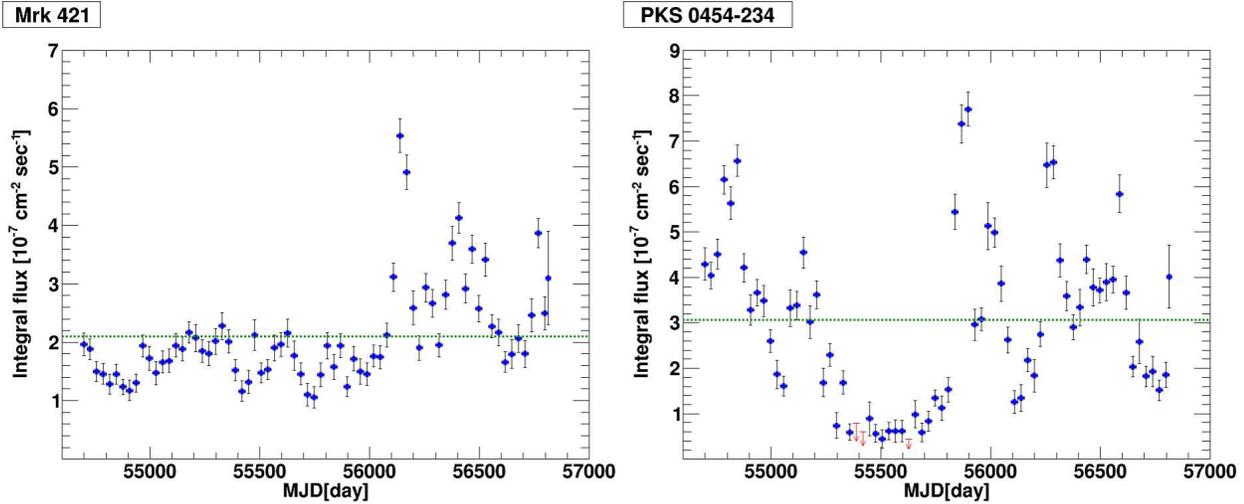}
\caption{Examples of the gamma-ray light curve about 6 years in 0.1--300 GeV. 
Blue plots: data points with a bin size of 30 days (bin size of last it is only about 4.4 days). 
Red allows: 95\% C.L. upper limits. 
Green dotted lines: average flux of whole period. 
Left panel: Mrk 421(HBL). 
Right panel: PKS 0454-234(FSRQ).}
\label{eq:LC}
\end{figure*}

We analysed the Fermi reprocessed pass 7 data between 2008 August 04 and 2014 June 09, using the unbinned likelihood analysis with the Fermi Science Tools package version v9r33p0 available from the Fermi Science Support Center (FSSC)\cite{FSSC}.
The likelihood analysis was selected that the events with photon energies in the range of 0.1-300 GeV and a Region Of Interest (ROI) of 10 degrees centered at the position of Table \ref{tab:blazarlist} sources. 
We used ``SOURCE'' class (``evclass = 2'') including both front and back events, because the ``SOURCE'' class is recommended for off-plane point source analysis by the likelihood analysis \cite{Anarecom}. 
We excluded events with zenith angles larger than 100 degrees and time intervals when the rocking angle was larger than 52 degrees. 
The set of the instrument response functions of ``P7REP\_SOURCE\_V15'' was applied. 
Models which were used in this study include the isotropic diffuse background (iso\_source\_v05.txt\cite{FSSCBG}), galactic diffuse background (gll\_iem\_v05\_rev1.fit\cite{FSSCBG}) and the
Second Fermi LAT Catalog (2FGL) sources in ROI of 10 degrees centered at the position of Table \ref{tab:blazarlist} sources. 
The spectrum model was according to the 2FGL. 
Target blazars (Table \ref{tab:blazarlist}) were fitted with a Log-Parabola (LP): 
\begin{equation}\label{eq:LogP}
\frac{dN}{dE} = N_0\left(\frac{\rm E}{{\rm E}_0}\right)
^{-(\alpha+\beta {\rm log}({\rm E}/{\rm E}_0))}
\end{equation}
because LP is typically used for modeling blazar spectra\cite{FermiSpectra}. 
Where $N_0$[cm$^{-2}$ s$^{-1}$ MeV$^{-1}$] is normalization parameter, E$_0$ [MeV] is scale parameter, $-(\alpha+\beta {\rm log}({\rm E}/{\rm E}_0))$ is spectrum index. 
If the parameter $\beta$ is zero, LP is equal to Powar-Law spectrum.
In this paper we fixed E$_0$ parameter of targets to E$_0$ = 100 MeV.

We judged the flux variation of target sources by some steps.
\\
Step I: gamma-ray light curves, which width of the time bins 
was fixed on 30 days (shortest bin in this study), was made.
\\
Step II: if calculated $F_{\rm var}$ was not required ``Selection Criteria'', we adopted more large bin size (60 days, 90 days, 150 days, and 300 days).
\\
``Selection Criteria'' were as follows, 
{\it I.} More than 40 \% of the calculated integral flux of each bin were detected.
{\it II.} Significant (over 2$\sigma$) variation was detected by the $\chi^2$ test in the analyzed period.
If the selection criteria {\it I.} and {\it II.} cleared, $F_{\rm var}$ of the target could be calculated.

\section{RESULTS and DISCUSSION}

\begin{figure*}[t]
\centering
\includegraphics[width=140mm]{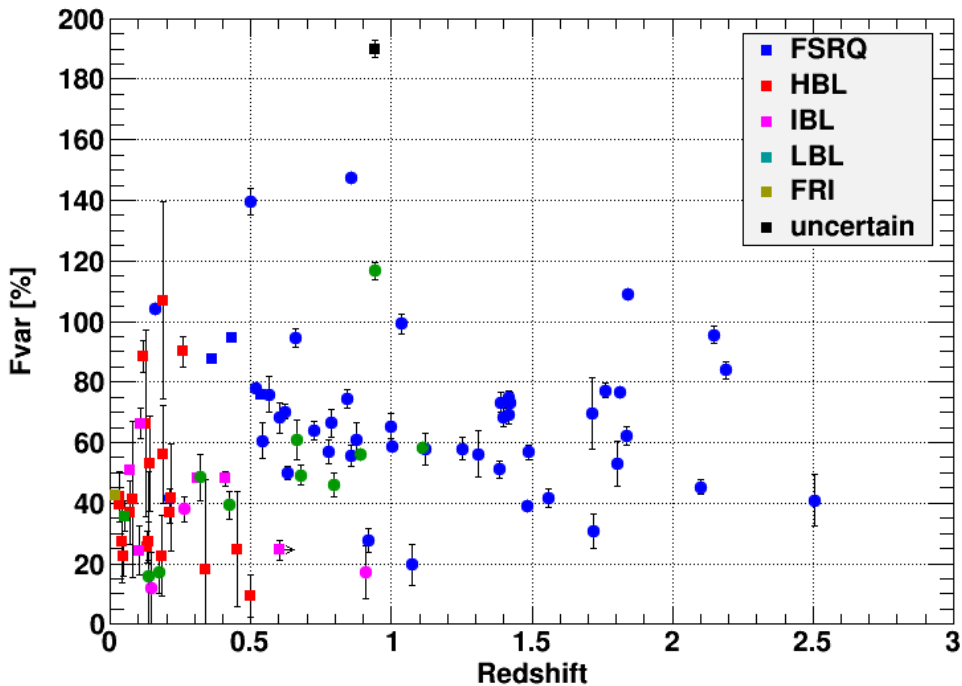}
\caption 
{The $F_{\rm var}$ as a function of the redshift.
Square and circle marks indicate TeVCat and not TeVCat sources,
respectively.
Each subclass of blazars are plotted in different colors
(Blue:FSRQ, Red: HBL, Magenta: IBL, Green: LBL, Yellow: FRI, Black: uncertain type.).} 
\label{fig:fvardist}
\end{figure*}

Figure \ref{eq:LC} shows the light curves of Mrk 421 and PKS 0454-234 as the light curve samples, which are the typical HBL and FSRQ sources, respectively.
The green dashed lines represent the average flux of whole period.
Each $F_{\rm var}$ and averaged flux were culculated as
  Mrk421:  $F_{\rm var}$ = 39.3 $\pm$ 1.3 \%   averaged flux = $(2.10 \pm 0.22) \times 10^{-7}$ cm$^{-1}$ s$^{-1}$, 
 PKS0454:  $F_{\rm var}$ =  58.7 $\pm$ 1.3 \%   averaged flux = $(3.07 \pm 0.32) \times 10^{-7}$ cm$^{-1}$ s$^{-1}$.
Note, the $F_{\rm var}$ calculation was performed only over the 9 TS bins.

Another $F_{\rm var}$s were obtained in the same method and 
the $F_{\rm var}$ as a function of the redshift is shown in Fig.\ref{fig:fvardist}.
Square and circle marks indicate TeVCat and not TeVCat sources,
respectively.
Each subclass of blazars are plotted in different colors
(Blue:FSRQ, Red: HBL, Magenta: IBL, Green: LBL, Yellow: FRI, Black: uncertain type.).
In Figure \ref{fig:fvardist}, 
Fvar indicates the variability amplitude 
of the GeV gamma-ray light curve.
GeV gamma ray from FSRQs could be detected at high redshift(z $>$ 0.5)
and have the large Fvar.
In addition, HBL and IBL assemble in z $<$ 0.5.
From these features, blazar subclass seems to change along the increasing redshift.\\
Peculiar features were as follows;\\
AO 0235+164 (z = 0.94, LBL) has particularly high $F_{\rm var}$ (117 $\pm$ 2.8) in LBLs.
This source was discussed that it might be FSRQ type blazar \cite{Ackermann2012,Stickel1994}; 
therefore, high $F_{\rm var}$ value of this source might be caused by the FSRQ like characters.
S3 0218+357 (z = 0.944, Uncertain type) which has the highest $F_{\rm var}$ (190 $\pm$ 2.8) in this analyzed sources is a gravitationally lensed blazar\cite{S3Atel}, 
hence the very high $F_{\rm var}$ value might be enhanced by the gravitationally lensed effect.

From Fig. \ref{fig:fvardist}, the $F_{\rm var}$ as a function of the redshift seems connection with FSRQs $\rightarrow$ LBLs $\rightarrow$ HBLs (IBLs).
This trend shows possibility of the activity evolution and blazar class evolution.

\begin{figure*}[t]
\centering
\includegraphics[width=135mm]{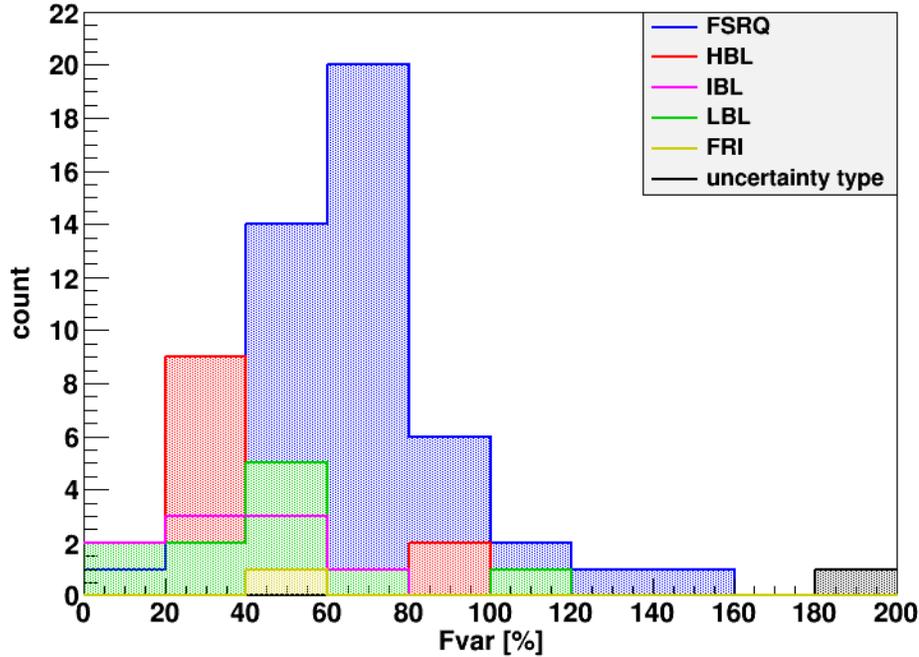}
\caption 
{The $F_{\rm var}$ distribution.
The width of $F_{\rm var}$ bin is 20\%.
The different colors show each subclass of blazars
(Blue: FSRQ, Red: HBL, Magenta: IBL, Green: LBL, Yellow: FRI, Black: uncertain type.).
} 
\label{fig:fvarhist}
\end{figure*}

Figure \ref{fig:fvarhist} shows the $F_{\rm var}$ histogram which is projected in the vertical axis of Fig. \ref{fig:fvardist}.
The different colors show each subclass of blazars
(Blue: FSRQ, Red: HBL, Magenta: IBL, Green: LBL, Yellow: FRI, Black: uncertain type).

However, there are some problems in this study.
First, the middle-high redshift (z $>$ 0.2) low $F_{\rm var}$ sources were not sufficient for discussions without selection effects.
Second, this study could not considered the short time scale variability ($<$ 30days).
Thus, it is necessary to analyze the low $F_{\rm var}$ sources and short time scale.

\newpage
\section{CONCLUSIONS}
We selected 102 AGNs (blazars) to reveal the relation 
between the evolution process of AGNs and the blazar sequence.
We applied the fractional variability
amplitude ($F_{\rm var}$) to calculate the variability amplitude considering the error.\\
\indent
The analyzed AGNs were selected from the second LAT AGN catalog (2LAC)\cite{2LAC} and TeVCat\cite{TeVCat}.
The analyzed data was Fermi reprocessed pass 7 data between 2008 August 04 and 2014 June 09, using the unbinned likelihood analysis with the Fermi Science Tools.\\
\indent
From these features, blazar subclass 
seems to change along the increasing redshift 
(connection with FSRQs $\rightarrow$ LBLs $\rightarrow$ HBLs (IBLs)).
This trend shows possibility of the activity evolution and blazar class evolution.
Peculiar features were as follows;
AO 0235+164 (z = 0.94, LBL) has particularly high $F_{\rm var}$ (117 $\pm$ 2.8) in LBLs.
This source was discussed that it might be FSRQ type blazar \cite{Ackermann2012,Stickel1994}; 
therefore, high $F_{\rm var}$ value of this source might be caused by the FSRQ like characters.
S3 0218+357 (z = 0.944, Uncertain type) which has the highest $F_{\rm var}$ (190 $\pm$ 2.8) in this analyzed sources is a gravitationally lensed blazar\cite{S3Atel}, 
hence the very high $F_{\rm var}$ value might be enhanced by the gravitationally lensed effect.
\\
\indent
However, there are some problems in this study.
First, the middle-high redshift (z $>$ 0.2) low $F_{\rm var}$ sources were not sufficient for discussions without selection effects.
Second, this study could not considered the short time scale variability ($<$ 30days).
Thus, It is necessary to analyze the low $F_{\rm var}$ sources and short time scale.



%

\bigskip 

\begin{thebibliography}{9}   




\bibitem{Ackermann2012}
M. Ackermann et al., ApJ, 751, (2012) 159

\bibitem{Elvis1992}
M. Elvis, et al., ApJS, 80, (1992) 257

\bibitem{2LAC}
Fermi-LAT Collaboration, ApJ, 743,(2011) 171

\bibitem{Fossati1998}
G. Fossati, et al., MNRAS, 299, (1998) 433

\bibitem{Furniss2013}
A. Furniss, ApJ, 768, (2013) 6

\bibitem{Kuhr1981}
H. K\"{u}r, et. al, A\&A, 45, (1981) 367

\bibitem{MAGIC2014}
MAGIC Collaboration, arXiv:1409.3389 (2014)    

\bibitem{Padovani1992}
P. Padovani and C.M. Urry, ApJ, 387, (1992) 449

\bibitem{Poutanen2008}
J. Poutanen, et. al, MNRAS, 1427, (2008) 389

\bibitem{Stickel1994}
M. Stickel, et al., A\&A Suppl. Ser., 105, (2014) 211 

\bibitem{Vaughan2003}
S. Vaughan, et. al, MNRAS, 345, (2003) 1271


\bibitem{TeVCat}
http://tevcat.uchicago.edu/ 

\bibitem{FSSC}
http://fermi.gsfc.nasa.gov/ssc/

\bibitem{FSSCBG}
http://fermi.gsfc.nasa.gov/ssc/data/access/lat/\\
BackgroundModels.html

\bibitem{Anarecom}
http://fermi.gsfc.nasa.gov/ssc/data/analysis/\\
documentation/Cicerone/Cicerone\_Data\_Exploration\\
/Data\_preparation.html

\bibitem{FermiSpectra}
http://fermi.gsfc.nasa.gov/ssc/data/analysis/\\
scitools/source\_models.html

\bibitem{S3Atel}
http://www.astronomerstelegram.org/?read=6349


\end{thebibliography}

\end{document}